%% file: creep.tex
\documentclass[prb,reprint,showpacs,amsmath,amssymb]{revtex4-1} %{revtex4-1}
\usepackage{bm}
\usepackage{graphicx}
\input{macros.tex}

\begin{document}
\title{Creep and fluidization in thermal amorphous solids}

\author{Samy Merabia, Fran\c{c}ois Detcheverry}

\affiliation{Univ Lyon, Universit\'e Claude Bernard Lyon 1, CNRS, Institut Lumi\`ere Mati\`ere, F-69622, VILLEURBANNE, France}

\begin{abstract}
When submitted to a constant mechanical load, 
many amorphous solids display power law creep followed by fluidization. 
A fundamental understanding of these processes is still far from being achieved. 
Here, we characterize creep and fluidization on the basis of a mesoscopic viscoplastic model 
that includes thermally activated yielding events 
and a broad distribution of energy barriers, 
which may be lowered under the effect of a local deformation. 
We relate the creep exponent observed  before fluidization 
to the  width of barrier distribution  
and to the specific form of stress redistribution following  yielding events. 
We show that Andrade creep is accompanied by local strain-hardening driven by stress redistribution
and find that the fluidization depends exponentially on the applied stress. 
The simulation results are interpreted in the light of  a mean-field analysis, 
and should help in rationalizing the creep phenomenology of amorphous solids.
\end{abstract}

\maketitle

\section{Introduction}
Creep is observed in a wide variety of materials  
including crystalline metals~\cite{prl_89-165501-2002}, soft crystals~\cite{prl_97-258303-2006}, 
polymeric and  metallic glasses~\cite{metals_3-77-2013,mm_41-4969-2008}, 
colloidal glasses~\cite{prl_108-255701-2012,sr_5-11884-2015} and gels~\cite{sm_6-3482-2010,sm_10-1555-2014}, 
and everyday complex fluids~\cite{ra_47-601-2008}. 
Typically, the strain first increases with time 
following a power law regime  often described as Andrade creep, 
with  $\epsilon(t) \sim t^p$ and a creep exponent~$p$  between~$0$ and~$1$. 
This creep regime is eventually interrupted by fluidization, 
after an elapsed time that decreases with the applied stress. 
Though the creep phenomenology is widespread, 
to date its understanding  remains only partial~\cite{arxiv_YSF-review-2016,cocis_19-549-2014}, 
in  particular for the underlying physical mechanism at play. 
Creep in metals is traditionally interpreted in terms of depinning and collective motion of 
dislocations~\cite{ijmr_100-1433-2009,prl_89-165501-2002}. 
No such framework exists for amorphous solids. 
%
%\com{ paper~\cite{prl_105-100601-2010}: c'est de la rupture a la fin non? donc c'est un peu different, 
%bitumen~\cite{pnas_102-10556-2009}, il y a de la fluidization a la fin?}

While molecular  simulations may provide a wealth of information 
on mechanical properties of disordered solids~\cite{msmse_19-083001-2011},  
the slow kinetics inherent to creep 
make it prohibitive to reach fluidization time.  
Following the pioneering work of Bulatov and Argon~\cite{msmse_2-167-1994}, 
mesoscopic models appear as an alternative to bridge both time and length scales 
between the molecular level and macroscopic, finite elements calculations~\cite{msmse_19-083001-2011}. 
The common idea is to coarse-grain fast microscopic motions, 
and retain only a minimal description of local plastic rearrangements or shear transformation zones (STZs), 
most importantly the long-range consequences of a single localized plastic event. 
Therefore, the essential ingredients  include a local yielding probability, 
and a spatially resolved dynamics for the stress redistribution, 
often described by an Eshelby form. 
While elasto-plastic models have now generated 
a sustained line of research~\cite{picard2004,prl_106-156001-2011,Nicolas2014,cocis_19-549-2014}, 
with much scrutiny on the the shear steady state,  
comparatively little attention has been devoted to analyze the creep 
dynamics in amorphous solids. 
Two noticeable exceptions are  
a spatially resolved Soft Glassy Rheology model~\cite{fielding2014} 
and a recent study by Bouttes and Vandembroucq~\cite{Bouttes2013},  which, however,
is restricted to logarithmic creep only.

The purpose of this letter is to propose an interpretation of creep 
on the basis of a mesoscopic model. 
We focus on thermal amorphous solids, such as polymeric  or metallic glasses, 
for which flow  proceeds through thermally activated localized events, 
and are characterized by a wide distribution of activation barriers. 
Using a  combination of numerical simulations and a mean-field analysis, 
we investigate how the creep exponent relates to the barrier distribution 
and show the role of the stress redistribution in the creep dynamics. 
Interestingly, the mesoscopic model reveals local strain-hardening during the creep regime: 
some regions may accumulate large levels of local stress, 
while others see their local stress decrease with the global shear rate. 
Such strain-hardening phenomenon, which differs from that seen in metals,
eventually triggers the fluidization of the material 
and allows to propose a simple law for the fluidization time. 
%

%the fluidization kinetics of the disordered solids. 
%Our conclusion  are based on simulations and mean-field analysis, 
%and will help to rationalize the timescales observed in the creep of thermal amorphous solids.
% The work is structured as follows: 
%we first present the mesoscopic model and its numerical implementation. 
%Next, we proceed to the mean-field analysis of the model to predict the creep exponent and strain-hardening. 
%The simulation results for creep are interpreted in the light of the mean field analysis, and we propose a simple law for the fluidization time. 
%We end up this letter by contrasting the creep mechanism that we unveil with the strain-hardening observed in crystalline metals. 

\section{Model} 
Our mesoscopic description relies on three main ingredients:
a distribution of yielding barriers, possibly modified by mechanical effect,  
thermal activation, 
and stress redistribution.
The system is divided into a collection of representative elements 
whose dimension corresponds to the size of a plastic event, 
and whose state is specified by an intrinsic energy barrier $E$ 
and  a local mechanical stress $\sigma$, assumed to be a scalar for simplicity. 
%\comSM{Here, we build on a scalar model for the sake of simplicity, but the generalization to a tensorial model is straightforward~\cite{Nicolas2014}.}
%
In a way similar to Eyring's model, 
thermal activation can trigger yielding with a rate
\be
\lambda(E,\sigma) = \taumicro^{-1} \exp \left[-\frac{E - h(\sigma)}{k_BT} \right]. 
\label{eq:lambda}
\ee
Here, $\taumicro$ is a microscopic time,  
$\kB$ is the Boltzmann constant and $T$ is the temperature. 
The function $h(\sigma)$ specifies how the  barrier may be lowered by the local stress; 
we will mainly consider a quadratic mechanical activation term 
% $h(\sigma) = \frac{\sigma^2 v_a} {4 \mu k_BT}$, 
$h(\sigma) = \sigma^2 v_a/(4 \mu k_BT)$,  
where $v_a$ is an activation volume, 
$\mu$ the infinite frequency shear modulus 
%and the factor $4$ is introduced so as to verify detailed balance condition
~\cite{Homer2010}. 
%but other choices will be briefly considered.
 After yielding, an element has his intrinsic barrier renewed from a distribution $\rhoE(E)$, 
its local stress put to zero, 
and the stress it carried is redistributed to other elements. 

Three types of stress  redistribution  are possible: 
``Eshelby'', ``mean-field'' and ``short-range''.
The former  originates in the Eshelby problem of an inclusion in an elastic matrix 
and its propagator  is quadrupolar~\cite{Eshelby1957,picard2004}
\be 
G_{ij}=\frac{2}{\pi r_{ij}^2} \cos (4 \theta_{ij})
\ee
where $G_{ij}$ is the contribution received by site~$i$ from a site~$j$, 
$r_{ij}=\vert \vc{r}_{ij} \vert$ is the distance between the two sites, 
and $\cos \theta_{ij}= (\vc{r}_{ij} \cdot \vc{e}_x) /r_{ij}$, where $\vc{e}_x$ is a unit vector along the direction of shear.
The mean-field propagator completely neglects spatial dependence and assign to all elements an identical contribution 
$G_{ij}=1/(N-1)$, $N$ being the total number of sites.   
To further assess the influence of the redistribution type, 
we will also consider a short-range propagator~\cite{acma_49-2017-2001}, 
for which the stress carried by an element is redistributed only to its nearest neighbours, 
as described in Ref.~\cite{Martens2012}. 

To fully specify the model,
it remains to choose the probability density of intrinsic barrier energy $\rhoE(E)$. 
In the following, we will concentrate mostly on a Gaussian distribution 
with mean  $\Ebar$  and variance $\sige^2$, 
\be
\rhoE(E) = \frac{1}{\sqrt{2 \pi \sige^2}} \exp \left[ -\frac{(E-\Ebar)^2}{2 {\sige}^2}\right].  
\label{eq:gaussian_DOS}
\ee
Such a Gaussian form is often assumed  in modelling  
the plastic behaviour  of polymer glasses~\cite{Hasan1995}, or more generally molecular glasses~\cite{Schirmacher2015,Xia2001}. 
Furthermore, it has been long recognized 
to be associated with a stretched exponential relaxation functions~\cite{Xia2001,Bouchaud1996}.
For the sake of analytical tractability, 
we will also  consider a barrier distribution which is 
exponential and has a width~$\alpha^{-1}$, 
\be
%\rhoE(E) =  \frac{1}{\sige} \exp \left(-\frac{(E-\Eo)}{\sige}\right) H[E-\Eo], 
\rhoE(E)   = \al  \exp \left[   - \al (E-\Eo) \right]   H[E-\Eo],          
\label{eq:exponential_DOS}
\ee
where $\Eo$ denotes the minimal energy barrier and $H$ is the Heaviside distribution, $H(x)=1$ if $x>0$ and $H(x)=0$ otherwise.   
The model is close to that of Ref.~\cite{Bouttes2013}, 
but the distribution of energy barriers has a width
which is finite rather than infinite, 
hence logarithmic creep is never observed. 

\begin{figure}[t]
\includegraphics[width=8.7cm]{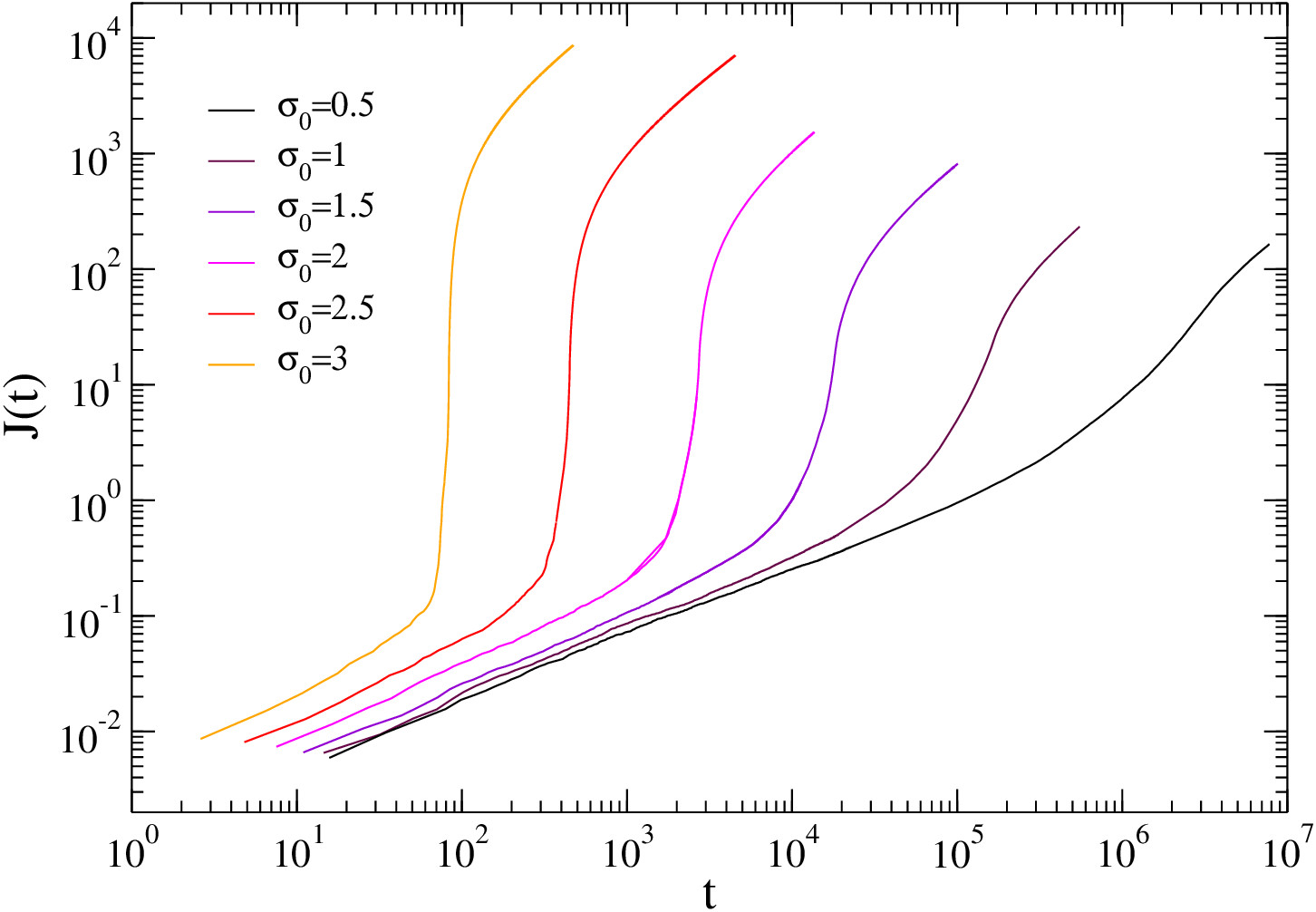}
\caption{ Creep compliance in simulations.   
The stress redistribution is of Eshelby type, 
and the distribution of energy barrier is Gaussian with $\sige=3$.  
Different values of applied stress $\sigo$ are shown.} 
\label{fig:creepcurve}
\end{figure}

\section{Simulations}
To solve  the model numerically, 
we discretize space with a two-dimensional square lattice and periodic boundary conditions. 
Initially,  each site~$i$ carries the same  stress~$\sigma_0$, 
and is assigned an energy barrier $E_i$ sampled from the steady distribution $\rho_E(E_i) \exp(E_i/k_BT)$. 
%The local energy barriers $E_i$ and stresses $\sigma_i$ are updated according to the local yielding rates $\lambda_i = \lambda(E_i,\sigma_i) = \taumicro^{-1} \exp \left[-\frac{E_i - h(\sigma_i)}{k_BT} \right]$. 
The creep process is simulated using  a Kinetic Monte Carlo~(KMC) algorithm~\cite{Bortz1975}. 
Given the yielding rates specified by Eq.~\eqref{eq:lambda} for all sites,  
each iteration selects a site~$i$  to yield, and generates a corresponding time increment. 
Upon yielding, the local stresses and the total strain  are updated as follows,
%which chooses at each iteration $n$ the site $i$ and the time increment $\Delta t_n=t_{n+1}-t_n$ according to the yielding rates $\lambda_i$. 
%If at a given iteration the site $i$ has been selected to yield, we update the values of the stresses and the energy barrier according to~:
\begin{eqnarray}
\label{eq:stress_dynamics}
\sigma_i \rightarrow 0, \:\:\:
\sigma_j \rightarrow \sigma_j + G_{ij} \sigma_i^{-} \; \mathrm{for} \; j \neq i, \quad
\epsilon \rightarrow \epsilon + \sigma_i^{-}/ 2\mu, 
\end{eqnarray} 
where $\sigma_i^{-}$ is the stress carried by the site $i$ prior to yielding. 
A new barrier energy is then chosen from the probability density~$\rho_E$. 
%The strain of the system $\epsilon(t)$ is also updated as 
% $\epsilon \rightarrow \epsilon + \sigma_i^{-}/ 2\mu$. 
%Simultaneously, we select a new energy barrier $E_i$ according to the density of states $\rho_E(E_i)$. 
We use a pseudo-spectral method  to carry out the elastic redistribution~\cite{picard2004}, 
%of Eq.~\ref{eq:stress_dynamics}
and  impose the sum rule $\forall i, \sum_{j \neq i} G_{ij}=1$, 
so that the  spatially averaged stress $\frac{1}{N} \sum_{i} \sigma_i=\sigma_0$ 
remains constant at all time, 
as required by the creep set-up.  
We have simulated systems with typical linear size $64$ and $256$, 
and verified that the results are not size dependent. 
In data presented below, 
we take $k_B T$ as energy unit, 
express stress in units where the shear modulus is $\mu=1$, 
and choose the time unit as $\taumicro \exp (\Ebar)$ or $\taumicro \exp (\Eo)$ 
in the Gaussian and exponential case respectively. 
Finally, the activation parameter is set to $v_a=1$. 
Once this choice is made, 
the only remaining parameter of the model is the distribution width $\sige$ or $\alpha^{-1}$. 
If not indicated otherwise, 
the stress redistribution is of Eshelby form, 
and $\rhoE$ is Gaussian.

%\comSM{In the paragraph below, the phenomenology is briefly described.}
Figure~\ref{fig:creepcurve} summarizes the creep phenomenology 
of our model. 
Whatever the applied stresses~$\sigo$, 
three regimes may be distinguished in 
the compliance curve $J(t)=\epsilon(t)/\sigo$. 
At early times, the compliance increases algebraically with time $J(t) \sim t^p$, 
where the creep exponent $p$ is found to be almost  independent of~$\sigo$. 
This first regime terminates with a  sharp increase in deformation, 
at a fluidization time~$t_{\rm f}$ 
that  decreases with the applied stress~$\sigo$. 
The system eventually settles into steady flow,   
where the strain increases linearly with time, $J(t) \sim t$. 
Below we investigate in turn the  primary creep and fluidization.  
To do so, 
we now develop a mean-field theory.

\section{Mean-field analysis}
When spatial dependence is entirely discarded,  
the system is completely described by  the probability density $P(E, \sigma, t)$ 
to find at time $t$  an element with energy barrier $E$ and subject to a stress $\sigma$. 
Our starting point is the evolution equation
%\begin{subequations}
%\label{eq:mf} 
\begin{align}
\partial_t P &= - \lambda(\Esig) P + \Y(t) \rhoE(E) \delta(\sigma) - S(t) \partial_\sigma P,   \label{eq:mfa} 
%   \Y(t)     &=  \avP{\lambda(\Esig)}, \label{eq:mfb}  \\ %\quad \:\:\:  S(t)=\avP{ \sigma  \lambda(\Esig)},   
%    S(t)     &=  \avP{ \sigma  \lambda(\Esig)}  \label{eq:mfc}
\end{align}
%\end{subequations}
where 
$\Y(t) =  \avP{\lambda(\Esig)}$, 
$S(t)  =  \avP{ \sigma  \lambda(\Esig)}$ 
and  $\avP{.}$ denotes an average over the full distribution $P(E, \sigma, t)$. 
In the RHS of Eq.~\eqref{eq:mfa},  
the first term originates from elements in  state ($\Esig$) that yield  with a rate~$\lambda(\Esig)$. 
$\Y(t)$ is the  average yielding rate, also called  material's fluidity~\cite{fielding2014}. 
Elements that have yielded  arrive in a renewed state 
with zero stress and an energy barrier randomly chosen in the distribution $\rhoE(E)$. 
The average rate of released stress $S(t)$ 
gathers the  contributions from all yielding elements, 
which is redistributed equally throughout the system,   
resulting in a drift term in $\sigma$ with velocity $S(t)$. 
One can check that the evolution equation implies two conserved quantities: 
the total probability and the total stress. 
While Eq.~\eqref{eq:mfa} may be written directly, 
a derivation is possible starting from 
a Boltzmann equation involving a stress collision operator.

The model defined  here  is related but distinct 
from the  Soft Glassy Rheology (SGR) model~\cite{Fielding2000}.  
With quadratic activation function $h(\sigma)\sim\sigma^2$ and an exponential $\rhoE(E)$, 
the model is formally equivalent to SGR  with  noise temperature $x=\alpha$.    
However, the interpretation is completely different, 
since as in the original trap model~\cite{Bouchaud1996}, 
$T$  here is really the temperature, 
not an effective noise resulting from yielding events elsewhere in the material. 
It was pointed out that the mechanical noise  in SGR should be 
``determined self-consistently by the interactions in the system''~\cite{pre_58-738-1998}. 
This key point is captured by the  redistribution term  $S(t) \partial_\sigma$, 
and is crucial for the creep situation,  a transient regime. 
In contrast to steady shear where the noise temperature is constant, 
the activity here is time-dependent, as it slowly declines during creep. 
Our description is also reminiscent of fiber bundles model (FBM) % \cite{pre_65-032502-2002}
but differs in an essential way~\cite{pre_65-032502-2002,rmp_82-499-2010}. 
In contrast to fibers that  permanently disappear once ruptured, 
elements that have yielded  are renewed and will again carry  a stress. 
The mean-field analysis  is used in the following  to provide a qualitative understanding ;  
to a large extent, it proves sufficient to rationalize what occurs in more realistic cases.  

\section{Creep regime}
We first consider the mean-field model 
and seek  the total strain  $\eps(t)=\int_0^t  S(t)/2\mu$ that follows a stress step. 
The model can be solved   if we neglect non-linear effects by setting $h(\sigma)=0$. 
In that case,  
the yielding of an element depends only on the time elapsed since its latest renewal 
and $\Y(t)$ can be computed without any reference to the local stress.  
The distribution of barriers is in  a steady  state  characterized by 
\be
\Pst(\tau) = \frac{\tau \rho(\tau)}{\taum}, \qquad \Yst = \frac{1}{\taum},
\ee
where, from now on, we use the intrinsic yielding time $\tau=e^E$ rather than the energy barrier,  
and $\av{.}$ denotes an average over the corresponding distribution~$\rho(\tau)$. 
The initial condition involves a uniform load on all elements 
and an equilibrated distribution of barriers, 
namely $P(\tau,\sigma,t=0) =  \Pst(\tau) \delta(\sigma-\sigo)$.  
We do not consider aging effects.

To solve the model, 
let's introduce $\sigb(\tau,t) = \int \dif\tau \sigma P(\tau,\sigma,t)$ which satisfies
\be
\partial_t\, \sigb =   -\frac{\sigb}{\tau} +  S(t) \Pst(\tau), \quad  S(t)= \int  \frac{d \tau}{\tau} \sigb(\tau,t). 
\label{eq:sigbt}
\ee
Using the condition $\int \dif t \: \sigb(\tau,t) = \sigo$ that holds at all time, 
and working with Laplace transforms, 
one obtains  the exact solution, valid for any distribution of barriers,
\be
\frac{S(s)}{\sigo} = \frac{1}{s R(s)} -1,  \quad R(s) =  \inttau \frac{\Pst(\tau)}{s+\tau^{-1}},         
\label{eq:Ss}
\ee
where $s$ is the Laplace variable and  indicates the nature of the function. 
Note that Eq.~\eqref{eq:Ss} can be rewritten as $\Jcomp(s) \Emod(s) = 1/s^2$,  
where $J(s)=\eps(s)/\sigo$ is the compliance, 
and $\Emod(s)=\mu R(s)$ is the relaxation modulus~\cite{book_flo-CreepRelaxViscoelastic}. 
The explicit expression
$\Emod(t)=\mu \int_0^{\infty} \Pst(\tau) e^{-t/\tau} \dif \tau$ 
 has a simple interpretation. 
The integral is the average fraction of elements that have never yielded  at time $t$, 
suggesting that sites that have already yielded at least once  
do not play any role, as if disappearing in FBM-like models. 
This interpretation is surprising at first sight
but understandable with the analysis of local stress presented below.

Though $R(s)$ can be obtained in closed form 
for some barrier distributions,   
taking the inverse Laplace transform of $1/R(s)$ proves impossible.  
Accordingly, we resort to a small-$s$ expansion 
and relying on Tauberian theorems~\cite{book_Feller-IntroProbTheo2}, 
we extract the asymptotic behavior of $S(t)$. 
For the sake of tractability, 
we consider an exponential distribution of barrier as defined in Eq.~\eqref{eq:exponential_DOS}, 
which translates into a power law distribution of yielding time 
$\rho(\tau) =  \al \tauo^ \al /\tau^{\al+1} H[\tau-\tauo]$, 
with $\tauo$ the minimum value. 
Assuming $\al >1$,  $R(s)$ can be expressed in terms of hypergeometric function as
$s R(s) =   {_2}F_1 \left( 1,-1+\al ,\al , -1/\tauo s \right)$.  
Several cases arise for the asymptotic behavior.
If $\al>2$,
then the  long-time behavior is Newtonian with 
$S(t) \sim \taum/\taus$, 
a result that holds more generally for any distribution $\rho(\tau)$ whose variance $\taus$ is finite. 
In the marginal case $\al=2$, one gets
$S(t) \sim  1/\ln(t/\tauo)$.  
More importantly, when $ 1<\al<2$, 
$S(t) \sim t^{\al-2}$, 
implying that the creep exponent is $p=\al-1$.  
Though the starting point given by Eq.~\eqref{eq:mfa} is different, 
those conclusions are in agreement with Ref.~\cite{Fielding2000}. 
We do not consider the case $\al<1$, 
as we  assume that the mean yielding $\taum$ is finite so that an equilibrated state exists.  
In the limit $\al \rightarrow 1^+$, the behavior approaches logarithmic creep, 
since $\taum$ grows without bound  and there are no more time scale in the system.

While those conclusions have been reached for an exponential $\rhoE(E)$, 
corresponding to a power law $\rho(\tau)$,  
 they are informative of other situations. 
First, if yielding times are bounded by a maximal value $\taumax$,
the long-time behavior is ultimately Newtonian but  up to $t \simeq \taumax$, 
we expect a transient regime similar to the asymptotic behavior described above. 
Second,  as soon as the distribution of energy is not narrowly peaked, 
there are widely different yielding times, 
and we expect that the creep exponent directly reflects the 
width of energy distribution.
\begin{figure}[t]
\includegraphics[width=8.5cm]{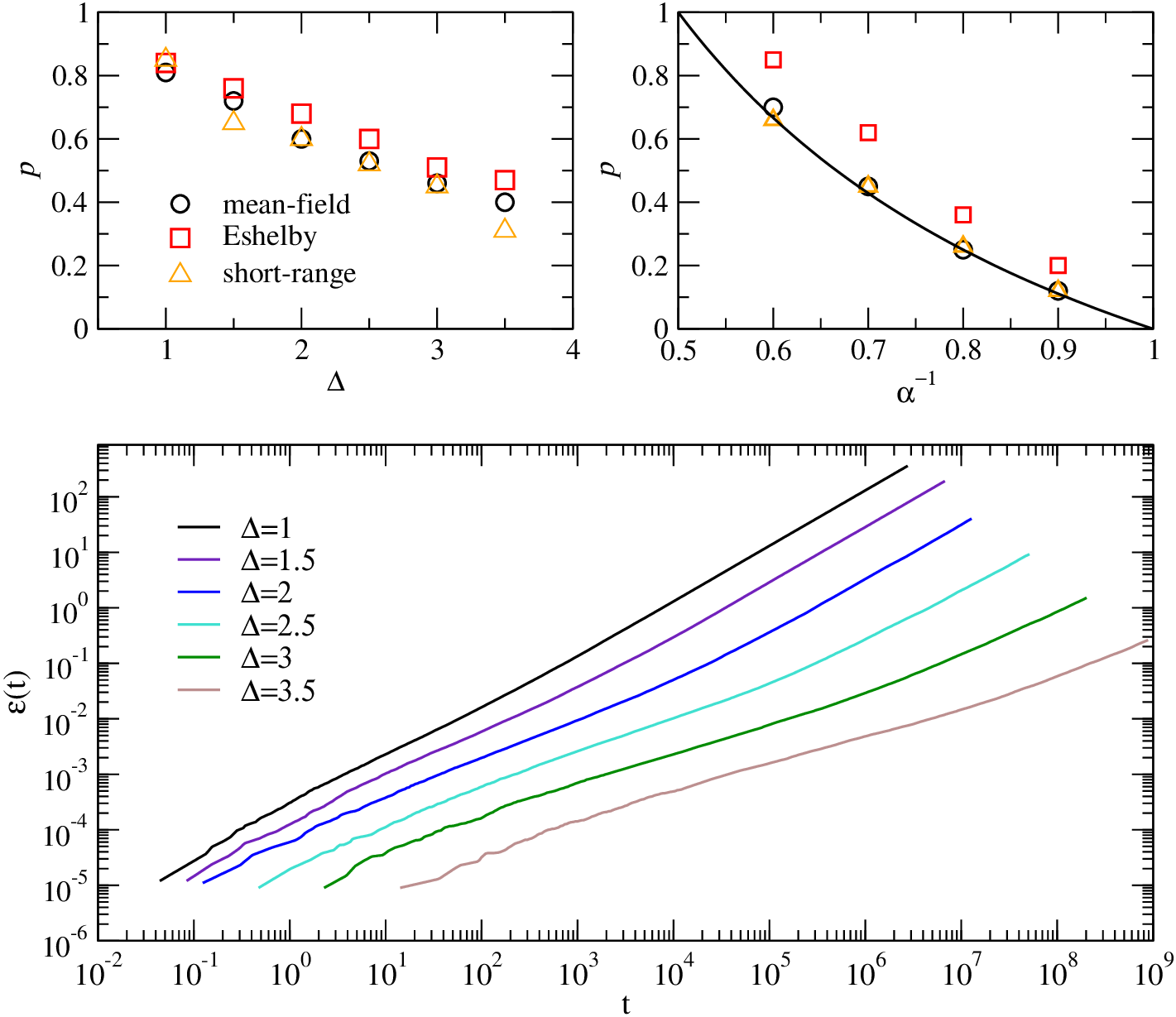}
\caption{
Creep regime and exponent. 
(Bottom) Strain curves with Eshelby redistribution 
and Gaussian $\rhoE$ of various  width~$\sige$.  
The applied stress is $\sigo= 0.01$. 
(Top) Creep exponent, as defined in the text, 
for three types of stress propagators, 
and for  Gaussian  and exponential $\rhoE$  (left and right respectively). 
In the latter, the line is the mean-field prediction. 
Symbol size is indicative of error bars.} 
\label{fig:creepexponent}
\end{figure}

With the mean-field prediction in hand, 
we now examine how the creep properties is affected by the type of stress redistribution 
and the choice of energy barrier. 
As a quantitative measure, we focus on the exponent characterizing the primary creep regime. 
In practice, a linear fit to $\epsilon(t)$ in bilogarithmic scale 
was used to get at all time an ``effective exponent'', 
the minimum of which is 
the creep exponent reported in Fig.~\ref{fig:creepexponent}. 
If $\rhoE$ is exponential, the asymptotic behavior is $\epsilon(t) \sim t^p$, 
and the minimum is attained in a plateau at the longest time. 
If $\rhoE$ is Gaussian, or bounded,  
then $\epsilon(t) = c t^p + t/\eta$, 
with $c$ a constant and $\eta$ the viscosity\footnote{Here the applied stress is so small that fluidization does not occur.}~\cite{Gueguen2015}.   
The effective exponent exhibits a minimum near the crossover between the two regimes, 
which, to be seen,  may require very long simulations, reaching up to $10^9$~KMC iterations.  
As shown in Fig.~\ref{fig:creepexponent} (top right),
the simulation data for the exponential $\rho_E$ is in full agreement with the mean-field prediction $p=\alpha-1$. 
Should we expect a similar result with Eshelby and short-range redistributions? 
On the one hand, 
given the long-range nature of elastic propagator, 
the mean-field theory could be expected to be exact~\cite{prl_102-175501-2009}. 
On the other hand, 
it was argued that the stress resulting from spatially distributed events 
is ``dominated by local contributions''~\cite{Fielding2009,picard2004}. 
Surprisingly, 
we find that within numerical accuracy, the mean-field and short-range exponent coincide, 
whereas the Eshelby case yields consistently higher values. 
This observation also applies to the Gaussian case. 
Overall,  we see  that the wider the barrier distribution, the slower the creep, 
but the value of creep exponent is sensitive to both the specific distribution of barriers  
and the form of the stress propagator.

\begin{figure}[t]
\center
\includegraphics[width=8.5cm]{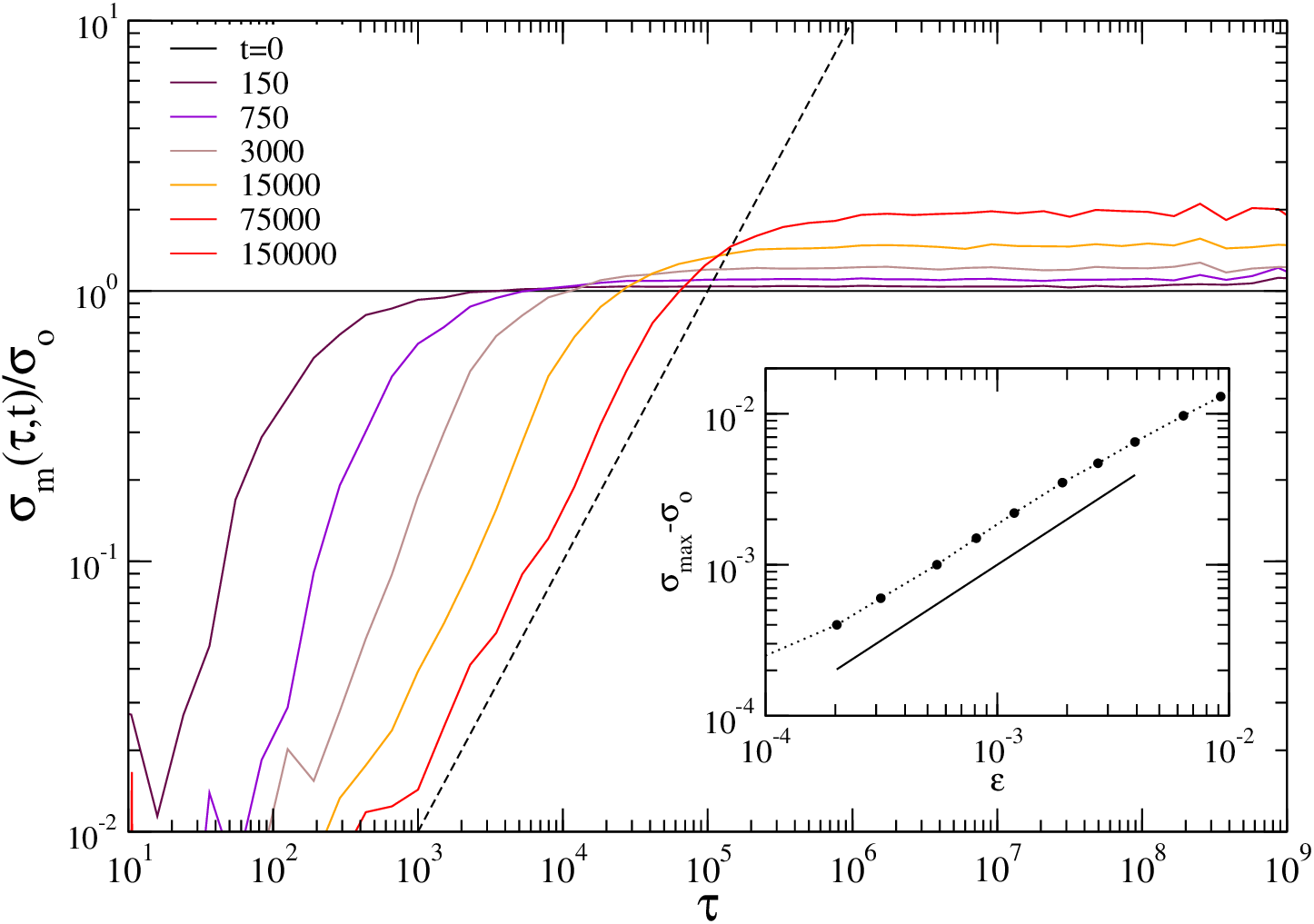}
\caption{
Analysis of local stress  during creep. 
The main graph shows how the  mean stress $\sigm(\tau ,t)$ 
carried by sites with  yielding time $\tau$ evolves in time.  
A plateau is seen at large~$\tau$.
(Inset) Plateau value, indicating the mean stress carried by ``slow'' elements, 
as a function of strain. 
The dashed and solid lines have slope unity. 
The simulation involves Eshelby redistribution, Gaussian $\rhoE$ with $\sige = 3$  and $\sigo = 0.01$.}
\label{fig:meanstress}
\end{figure}

\section{Local stress}
%\com{Les deux paragraphes suivants sont redondants: les fondre en un seul ?}
%\comSM{En même temps, je trouve c'est bien comme ca}
%
To get further insight in the mesoscopic dynamics during the creep regime,
we have conducted an analysis of the local stress carried by elements.  
Of particular attention is the relation between the local stress $\sigma_i$ carried by an element~$i$, 
and its instantaneous yielding time $\tau_i$. 
% Since we are first examining the primary creep regime, this yielding time does not depend 
% on the stress $\sigma_i$. 
% FFF je ne comprends pas la phrase ci-dessus. 
Figure~\ref{fig:meanstress} reveals a strongly heterogeneous dynamics during primary creep. 
Indeed, ``fast'' elements
having a low energy barrier carry on average a small amount of stress $\sigma_i \ll \sigo$, 
while ``slow'' elements support most of the stress. 
Noticeably, the level of stress borne by these elements increases with time. 
As shown in the inset of Fig.~\ref{fig:meanstress}, 
this increase is approximately proportional to the local strain~$\epsilon(t)$. 
%\comSM{A enlever je pense, car on developpe pas assez: This strain-hardening behavior is reminiscent of the dislocation dynamics in metals.}   
Such strain-hardening, 
here understood as an increase in local stress required to produce additional strain, 
may be simply explained in the framework of the mean-field analysis presented above. 
Using Eqs.~\eqref{eq:sigbt}-\eqref{eq:Ss}, 
one gets for  $\sigm(\tau,t)$, the mean stress carried at time~$t$ by elements with yield time~$\tau$, 
\be
\sigm(\tau,s) = \frac{ \sigb(\tau,s)}{ \Pst(\tau)} = \frac{\sigo+S(s)}{s+\tau^{-1}}. 
\ee
and obtain in the two limits, 
\begin{subequations}
\begin{align}
\tau \ll t, \qquad   &\sigm(\tau,t)=\mu \tau \epsdot(t),          \label{eq:popa}    \\
\tau \gg t, \qquad   &\sigm(\tau,t)=\mu  \eps(t) + \sigo.     \label{eq:popb}
\end{align}               
\end{subequations}
Schematically, one can identify two populations of sites, 
respectively fast and slow depending on the value of the 
local yielding time~$\tau$ as compared to the elapsed time~$t$. 
On the one hand, the sites that have yielded already 
and that are carrying a stress decreasing in time as $\epsdot(t)$. 
On the other hand, the resistant sites that have not yielded yet, 
and who carry a stress increasing as $\eps(t)$.  
%in agreement with the simulation results.  
In Fig.~\ref{fig:meanstress}, 
one sees that Eqs.~\eqref{eq:popa} and~\eqref{eq:popb} 
apply to a good approximation,  
even though the propagator is of Eshelby type rather than mean-field.

\section{Fluidization time}
\begin{figure}[t]
\center
\includegraphics[width=8cm]{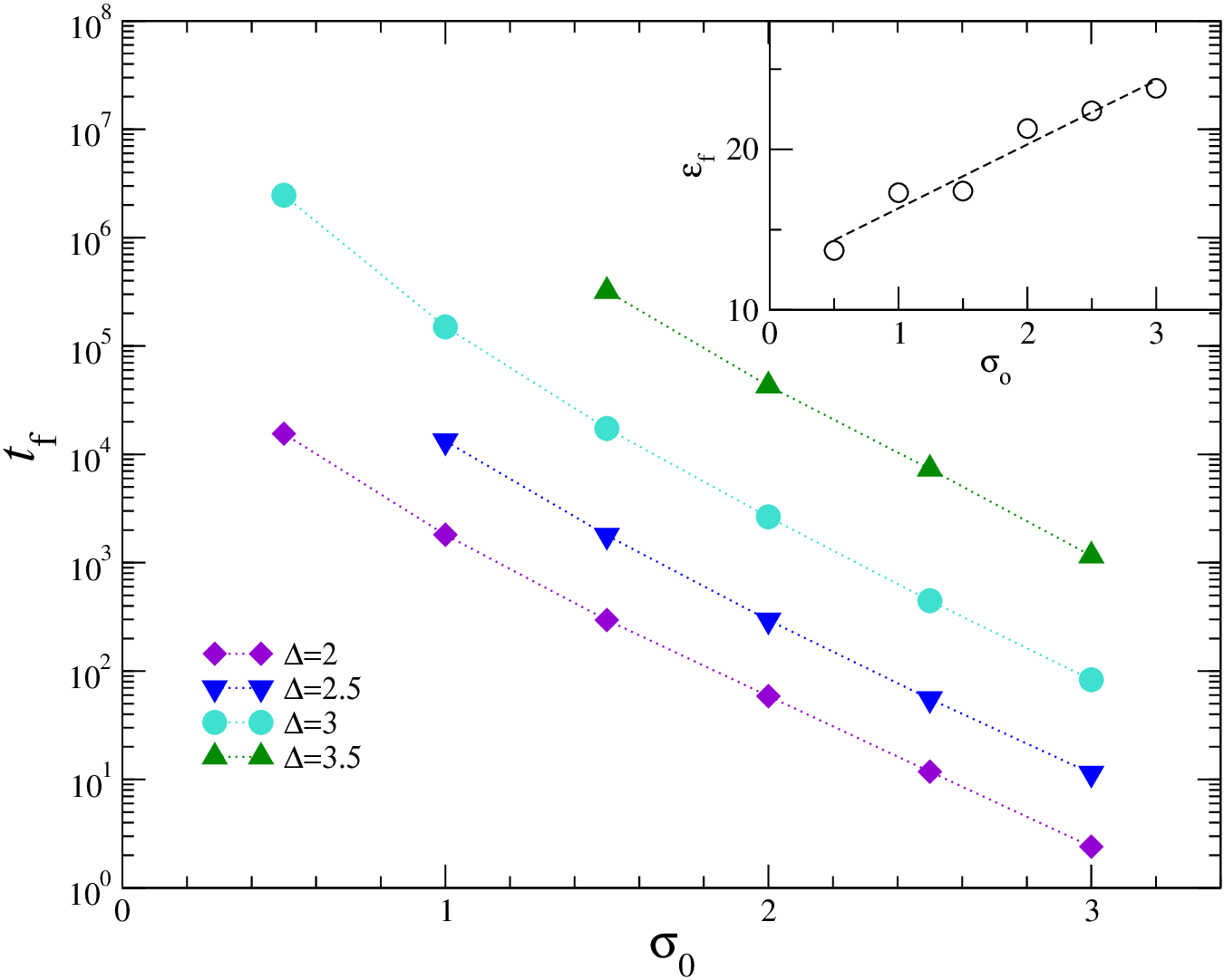}
\caption{
Fluidization time~$\tf$ as a function of the applied stress~$\sigo$. 
(Inset) 
Strain at fluidization $\epsf$ as a function of $\sigo$. 
The dashed line is a linear fit.
Here,  redistribution is of Eshelby type, 
$\rhoE$ is Gaussian with  width~$\sige$. }
\label{fig:tf}
\end{figure}

At the fluidization transition, the deformation increases sharply, 
and we have observed strain localization, 
as already noticed in Ref.~\cite{fielding2014}. 
In particular, the standard deviation of the local strain goes through a maximum, 
which is used to pinpoint the fluidization time~$\tf$. 
Figure~\ref{fig:tf} reveals an exponential dependence of the fluidization time~$\tf(\sigo)$ 
on the applied stress. 

To  rationalize this behavior, 
we make use of two observations. 
First, the strain at fluidization $\epsf=\eps(\tf)$ varies only weakly with $\sigo$, 
namely $\epsf(\sigo) \approx \tilde{\epsf} -\zeta \sigo$, 
%\footnote{We have found that the factor $\zeta$ can be negative or positive depending on the case considered.}, 
as previously observed in some experiments~\cite{prl_97-258303-2006,ra_47-601-2008}. 
Second, the particular form of $h(\sigma)$ does not appear to have the leading role 
in the $\tf(\sigo)$ relation since we also found an exponential dependence
when $h(\sigma)$ is linear rather than quadratic. 
%\footnote{Do we need more simulations to ensure this point?}. 
%
Consider the plateau in $\sigma$ associated to slow sites, 
which at a time~$t$, ranges from~$t$ to~$\taumax$, the largest relaxation time in the system. 
We reintroduce the effect of activation in an approximate manner, 
with $\sigma(t)$ estimated from the  solution with no activation term ($h=0$) found above,  
thus leading to a shift factor $\exp\left[ -h(\sigm(\tau,t)) \right]$ 
for an element with intrinsic time~$\tau$. 
Now, we postulate that the fluidization occurs where there is no more element 
whose actual relaxation time is longer than the elapsed time, 
\be
\tf = \taumax \exp\left[ -h(\sigo+ \mu \epsf) \right],  
\ee
that is, activation effects have shifted the longest intrinsic relaxation time to a value below~$\tf$. 
Assuming $\epsf$ is strictly independent of $\sigo$, 
and expanding at first order in $\sigo \ll \mu \epsf$, one gets
\be
\tf = C \exp\left[ -\sigo/\sigmat \right], 
\ee
with $C=\taumax \exp\left[  -h( \mu \epsf) \right]$ and $\sigmat = 1/h'(\mu \epsf)$. 
A similar argument applies if $\epsf(\sigo)$ exhibits a linear dependence as considered above. 
As regards the dependence in the width~$\sige$ of barrier distribution,  
we note that the prefactor~$C$ may change significantly, 
as $\taumax$ increases with~$\sige$. 
%On the other hand, $\sigmat$, which governs the slope, 
%varies very little with~$\sige$, as seen in Fig.~\ref{fig:tf}.  
%\com{This suggests that $\epsf$ is almost independent of $\sige$ as well, as indeed seen in simulations}
%
In experiments, alongside power law dependence for carbopol gels~\cite{prl_104-208301-2010,sm_8-4151-2012}, 
an exponential $\tf(\sigo)$ was reported in carbon black gels~\cite{sm_10-1555-2014,sm_6-3482-2010}, 
thermo-reversible silica gels~\cite{jrheo_51-623-2007}
and protein gels~\cite{sm_8-3657-2012}.  
Within our mesoscopic model, this phenomenology can be attributed 
to activated dynamics with energy barriers that are lowered by the applied stress. 

%In contrast, power law behaviours have been experimentally reported for carbopol gels~\cite{prl_104-208301-2010,sm_8-4151-2012}, 
%and yet there does not exist any model that explains this type of fluidization kinetics. While most of the systems cited above are yield stress fluids, one should keep in mind that our model is not designed to describe a yield stress fluid, as the steady rheology is best described by 
%a power law fluid.}

Before concluding, we  briefly comment on the steady state reached after fluidization, 
that is characterized by a constant shear rate $\dot{\gamma}$, 
as observed in colloidal glasses~\cite{prl_108-255701-2012}. 
Simulations show that the final state attained during steady creep 
is identical to that reached upon constant deformation rate $\dot{\gamma}$. 
For exponential barrier distribution, 
the flow curve indicates a power law fluid $\sigma \sim \dot{\gamma}^{\alpha}$, 
in agreement with a mean-field analysis. 
In the Gaussian case, 
one finds a logarithmic behavior $\sigma \sim \ln (\dot{\gamma})$.

%In agreement with the mean field analysis, the steady state rheology is described either by a power law fluid $\sigma \sim \dot{\gamma}^{\alpha}$
%or by a logarithmic behavior $\sigma \sim \ln (\dot{\gamma})$ depending on the form of the energy barrier distribution.

\section{Conclusion}
Through the consideration of a mesoscopic viscoplastic model, 
we demonstrated that the creep dynamics is directly related 
to distribution of  energy barriers, 
and to the form of the stress redistribution subsequent to yielding. 
Moreover, our simulations show that primary creep regime is accompanied 
by local strain-hardening, resulting from the existence of a broad distribution of yielding times. 
Strain hardening is also key to understand the fluidization process, which here displays an exponential dependence on the applied stress, as seen in experiments on colloidal gels.       
We have focused on amorphous solids where thermally activated yielding events 
play the leading role. 
Creep and fluidization are also observed in athermal systems such as carbopol gels~\cite{prl_104-208301-2010,sm_8-4151-2012}, 
which are yield stress fluids. 
It remains to address this important class of materials.

\section{Acknowledgements}
We are grateful to C.~Barentin, M.~Le~Merrer and L.~Vanel, 
as well as T.~Divoux and S.~Manneville, 
for introducing us to the phenomenology of creep in soft materials 
and for stimulating discussions. 
Part of the simulations have been run at PSMN, P\^ole Scientifique de Mod\'elisation
Num\'erique, Lyon.

\bibliographystyle{eplbib.bst}
\bibliography{creep,shortbib}

\end{document}

%% file: macros.tex
\usepackage{color}

\newcommand{\be}{\begin{eqnarray}}
\newcommand{\ee}{\end{eqnarray}}

\newcommand{\al}{\alpha}
\newcommand{\dif}{\mathrm{d}}
\newcommand{\eps}{\epsilon}
\newcommand{\epsdot}{\dot{\epsilon}}
\newcommand{\epsf}{\epsilon_{\mathrm{f}}}
\newcommand{\Emod}{G}
\newcommand{\Ebar}{\bar{E}}
\newcommand{\Eo}{E_o}
\newcommand{\Esig}{E,\sigma}

\newcommand{\inttau}{\int \dif \tau}
\newcommand{\Jcomp}{J}
\newcommand{\kB}{k_B}

\newcommand{\Pst}{P_{\mathrm{st}}}
\newcommand{\rhoE}{\rho_{\mathrm{E}}}
\newcommand{\sigb}{\overline{\sigma}}
\newcommand{\sige}{\varDelta} %\sigma_E
\newcommand{\sigm}{\sigma_{\mathrm{m}}}
\newcommand{\sigmat}{\tilde{\sigma}}
\newcommand{\sigo}{\sigma_{\mathrm{o}}}

\newcommand{\taumax}{\tau_{\mathrm{max}}}
\newcommand{\taumicro}{\tau_{\mathrm{m}}}
\newcommand{\tf}{t_{\mathrm{f}}}
\newcommand{\Y}{Y}
\newcommand{\Yst}{Y_{\mathrm{st}}}

\newcommand{\av}[1]{\langle #1 \rangle}  %_{\rho}
\newcommand{\avP}[1]{\langle #1 \rangle_P}  %_{\rho}
\newcommand{\vc}[1]{\mathbf{#1}}
\newcommand{\taum}{\av{\tau}}
\newcommand{\taus}{\av{\tau^2}}
\newcommand{\tauo}{\tau_o}

%% file: creep.bbl
\begin{thebibliography}{41}%
\makeatletter
\providecommand \@ifxundefined [1]{%
 \@ifx{#1\undefined}
}%
\providecommand \@ifnum [1]{%
 \ifnum #1\expandafter \@firstoftwo
 \else \expandafter \@secondoftwo
 \fi
}%
\providecommand \@ifx [1]{%
 \ifx #1\expandafter \@firstoftwo
 \else \expandafter \@secondoftwo
 \fi
}%
\providecommand \natexlab [1]{#1}%
\providecommand \enquote  [1]{``#1''}%
\providecommand \bibnamefont  [1]{#1}%
\providecommand \bibfnamefont [1]{#1}%
\providecommand \citenamefont [1]{#1}%
\providecommand \href@noop [0]{\@secondoftwo}%
\providecommand \href [0]{\begingroup \@sanitize@url \@href}%
\providecommand \@href[1]{\@@startlink{#1}\@@href}%
\providecommand \@@href[1]{\endgroup#1\@@endlink}%
\providecommand \@sanitize@url [0]{\catcode `\\12\catcode `\$12\catcode
  `\&12\catcode `\#12\catcode `\^12\catcode `\_12\catcode `\%12\relax}%
\providecommand \@@startlink[1]{}%
\providecommand \@@endlink[0]{}%
\providecommand \url  [0]{\begingroup\@sanitize@url \@url }%
\providecommand \@url [1]{\endgroup\@href {#1}{\urlprefix }}%
\providecommand \urlprefix  [0]{URL }%
\providecommand \Eprint [0]{\href }%
\providecommand \doibase [0]{http://dx.doi.org/}%
\providecommand \selectlanguage [0]{\@gobble}%
\providecommand \bibinfo  [0]{\@secondoftwo}%
\providecommand \bibfield  [0]{\@secondoftwo}%
\providecommand \translation [1]{[#1]}%
\providecommand \BibitemOpen [0]{}%
\providecommand \bibitemStop [0]{}%
\providecommand \bibitemNoStop [0]{.\EOS\space}%
\providecommand \EOS [0]{\spacefactor3000\relax}%
\providecommand \BibitemShut  [1]{\csname bibitem#1\endcsname}%
\let\auto@bib@innerbib\@empty
%</preamble>
\bibitem [{\citenamefont {Miguel}\ \emph {et~al.}(2002)\citenamefont {Miguel},
  \citenamefont {Vespignani}, \citenamefont {Zaiser},\ and\ \citenamefont
  {Zapperi}}]{prl_89-165501-2002}%
  \BibitemOpen
  \bibfield  {author} {\bibinfo {author} {\bibfnamefont {M.-C.}\ \bibnamefont
  {Miguel}}, \bibinfo {author} {\bibfnamefont {A.}~\bibnamefont {Vespignani}},
  \bibinfo {author} {\bibfnamefont {M.}~\bibnamefont {Zaiser}}, \ and\ \bibinfo
  {author} {\bibfnamefont {S.}~\bibnamefont {Zapperi}},\ }\href@noop {}
  {\bibfield  {journal} {\bibinfo  {journal} {Phys. Rev. Lett.}\ }\textbf
  {\bibinfo {volume} {89}},\ \bibinfo {pages} {165501} (\bibinfo {year}
  {2002})}\BibitemShut {NoStop}%
\bibitem [{\citenamefont {Bauer}\ \emph {et~al.}(2006)\citenamefont {Bauer},
  \citenamefont {Oberdisse},\ and\ \citenamefont {Ramos}}]{prl_97-258303-2006}%
  \BibitemOpen
  \bibfield  {author} {\bibinfo {author} {\bibfnamefont {T.}~\bibnamefont
  {Bauer}}, \bibinfo {author} {\bibfnamefont {J.}~\bibnamefont {Oberdisse}}, \
  and\ \bibinfo {author} {\bibfnamefont {L.}~\bibnamefont {Ramos}},\
  }\href@noop {} {\bibfield  {journal} {\bibinfo  {journal} {Phys. Rev. Lett.}\
  }\textbf {\bibinfo {volume} {97}},\ \bibinfo {pages} {258303} (\bibinfo
  {year} {2006})}\BibitemShut {NoStop}%
\bibitem [{\citenamefont {Egami}\ \emph {et~al.}(2013)\citenamefont {Egami},
  \citenamefont {Iwashita},\ and\ \citenamefont {Dmowski}}]{metals_3-77-2013}%
  \BibitemOpen
  \bibfield  {author} {\bibinfo {author} {\bibfnamefont {T.}~\bibnamefont
  {Egami}}, \bibinfo {author} {\bibfnamefont {T.}~\bibnamefont {Iwashita}}, \
  and\ \bibinfo {author} {\bibfnamefont {W.}~\bibnamefont {Dmowski}},\
  }\href@noop {} {\bibfield  {journal} {\bibinfo  {journal} {Metals}\ }\textbf
  {\bibinfo {volume} {3}},\ \bibinfo {pages} {77} (\bibinfo {year}
  {2013})}\BibitemShut {NoStop}%
\bibitem [{\citenamefont {Riggleman}\ \emph {et~al.}(2008)\citenamefont
  {Riggleman}, \citenamefont {Schweizer},\ and\ \citenamefont {{De
  Pablo}}}]{mm_41-4969-2008}%
  \BibitemOpen
  \bibfield  {author} {\bibinfo {author} {\bibfnamefont {R.~A.}\ \bibnamefont
  {Riggleman}}, \bibinfo {author} {\bibfnamefont {K.~S.}\ \bibnamefont
  {Schweizer}}, \ and\ \bibinfo {author} {\bibfnamefont {J.~J.}\ \bibnamefont
  {{De Pablo}}},\ }\href@noop {} {\bibfield  {journal} {\bibinfo  {journal}
  {Macromolecules}\ }\textbf {\bibinfo {volume} {41}},\ \bibinfo {pages} {4969}
  (\bibinfo {year} {2008})}\BibitemShut {NoStop}%
\bibitem [{\citenamefont {Siebenb{\"{u}}rger}\ \emph
  {et~al.}(2012)\citenamefont {Siebenb{\"{u}}rger}, \citenamefont {Ballauff},\
  and\ \citenamefont {Voigtmann}}]{prl_108-255701-2012}%
  \BibitemOpen
  \bibfield  {author} {\bibinfo {author} {\bibfnamefont {M.}~\bibnamefont
  {Siebenb{\"{u}}rger}}, \bibinfo {author} {\bibfnamefont {M.}~\bibnamefont
  {Ballauff}}, \ and\ \bibinfo {author} {\bibfnamefont {T.}~\bibnamefont
  {Voigtmann}},\ }\href@noop {} {\bibfield  {journal} {\bibinfo  {journal}
  {Phys. Rev. Lett.}\ }\textbf {\bibinfo {volume} {108}},\ \bibinfo {pages}
  {255701} (\bibinfo {year} {2012})}\BibitemShut {NoStop}%
\bibitem [{\citenamefont {Sentjabrskaja}\ \emph {et~al.}(2015)\citenamefont
  {Sentjabrskaja}, \citenamefont {Chaudhuri}, \citenamefont {Hermes},
  \citenamefont {Poon}, \citenamefont {Horbach}, \citenamefont {Egelhaaf},\
  and\ \citenamefont {Laurati}}]{sr_5-11884-2015}%
  \BibitemOpen
  \bibfield  {author} {\bibinfo {author} {\bibfnamefont {T.}~\bibnamefont
  {Sentjabrskaja}}, \bibinfo {author} {\bibfnamefont {P.}~\bibnamefont
  {Chaudhuri}}, \bibinfo {author} {\bibfnamefont {M.}~\bibnamefont {Hermes}},
  \bibinfo {author} {\bibfnamefont {W.~C.~K.}\ \bibnamefont {Poon}}, \bibinfo
  {author} {\bibfnamefont {J.}~\bibnamefont {Horbach}}, \bibinfo {author}
  {\bibfnamefont {S.~U.}\ \bibnamefont {Egelhaaf}}, \ and\ \bibinfo {author}
  {\bibfnamefont {M.}~\bibnamefont {Laurati}},\ }\href@noop {} {\bibfield
  {journal} {\bibinfo  {journal} {Sci. Rep.}\ }\textbf {\bibinfo {volume}
  {5}},\ \bibinfo {pages} {11884} (\bibinfo {year} {2015})}\BibitemShut
  {NoStop}%
\bibitem [{\citenamefont {Gibaud}\ \emph {et~al.}(2010)\citenamefont {Gibaud},
  \citenamefont {Frelat},\ and\ \citenamefont {Manneville}}]{sm_6-3482-2010}%
  \BibitemOpen
  \bibfield  {author} {\bibinfo {author} {\bibfnamefont {T.}~\bibnamefont
  {Gibaud}}, \bibinfo {author} {\bibfnamefont {D.}~\bibnamefont {Frelat}}, \
  and\ \bibinfo {author} {\bibfnamefont {S.}~\bibnamefont {Manneville}},\
  }\href@noop {} {\bibfield  {journal} {\bibinfo  {journal} {Soft Matter}\
  }\textbf {\bibinfo {volume} {6}},\ \bibinfo {pages} {3482} (\bibinfo {year}
  {2010})}\BibitemShut {NoStop}%
\bibitem [{\citenamefont {Grenard}\ \emph {et~al.}(2014)\citenamefont
  {Grenard}, \citenamefont {Divoux}, \citenamefont {Taberlet},\ and\
  \citenamefont {Manneville}}]{sm_10-1555-2014}%
  \BibitemOpen
  \bibfield  {author} {\bibinfo {author} {\bibfnamefont {V.}~\bibnamefont
  {Grenard}}, \bibinfo {author} {\bibfnamefont {T.}~\bibnamefont {Divoux}},
  \bibinfo {author} {\bibfnamefont {N.}~\bibnamefont {Taberlet}}, \ and\
  \bibinfo {author} {\bibfnamefont {S.}~\bibnamefont {Manneville}},\
  }\href@noop {} {\bibfield  {journal} {\bibinfo  {journal} {Soft Matter}\
  }\textbf {\bibinfo {volume} {10}},\ \bibinfo {pages} {1555} (\bibinfo {year}
  {2014})},\ \Eprint {http://arxiv.org/abs/1310.0385} {arXiv:1310.0385}
  \BibitemShut {NoStop}%
\bibitem [{\citenamefont {Caton}\ and\ \citenamefont
  {Baravian}(2008)}]{ra_47-601-2008}%
  \BibitemOpen
  \bibfield  {author} {\bibinfo {author} {\bibfnamefont {F.}~\bibnamefont
  {Caton}}\ and\ \bibinfo {author} {\bibfnamefont {C.}~\bibnamefont
  {Baravian}},\ }\href@noop {} {\bibfield  {journal} {\bibinfo  {journal}
  {Rheol. Acta}\ }\textbf {\bibinfo {volume} {47}},\ \bibinfo {pages} {601}
  (\bibinfo {year} {2008})}\BibitemShut {NoStop}%
\bibitem [{\citenamefont {Bonn}\ \emph {et~al.}(2015)\citenamefont {Bonn},
  \citenamefont {Paredes}, \citenamefont {Denn}, \citenamefont {Berthier},
  \citenamefont {Divoux},\ and\ \citenamefont
  {Manneville}}]{arxiv_YSF-review-2016}%
  \BibitemOpen
  \bibfield  {author} {\bibinfo {author} {\bibfnamefont {D.}~\bibnamefont
  {Bonn}}, \bibinfo {author} {\bibfnamefont {J.}~\bibnamefont {Paredes}},
  \bibinfo {author} {\bibfnamefont {M.~M.}\ \bibnamefont {Denn}}, \bibinfo
  {author} {\bibfnamefont {L.}~\bibnamefont {Berthier}}, \bibinfo {author}
  {\bibfnamefont {T.}~\bibnamefont {Divoux}}, \ and\ \bibinfo {author}
  {\bibfnamefont {S.}~\bibnamefont {Manneville}},\ }\href@noop {} {\bibfield
  {journal} {\bibinfo  {journal} {arXiv}\ ,\ \bibinfo {pages} {1502.05281}}
  (\bibinfo {year} {2015})},\ \Eprint {http://arxiv.org/abs/1502.05281}
  {arXiv:1502.05281} \BibitemShut {NoStop}%
\bibitem [{\citenamefont {Voigtmann}(2014)}]{cocis_19-549-2014}%
  \BibitemOpen
  \bibfield  {author} {\bibinfo {author} {\bibfnamefont {T.}~\bibnamefont
  {Voigtmann}},\ }\href@noop {} {\bibfield  {journal} {\bibinfo  {journal}
  {Curr. Opin. Colloid Interface Sci.}\ }\textbf {\bibinfo {volume} {19}},\
  \bibinfo {pages} {549} (\bibinfo {year} {2014})}\BibitemShut {NoStop}%
\bibitem [{\citenamefont {Louchet}\ and\ \citenamefont
  {Duval}(2009)}]{ijmr_100-1433-2009}%
  \BibitemOpen
  \bibfield  {author} {\bibinfo {author} {\bibfnamefont {F.}~\bibnamefont
  {Louchet}}\ and\ \bibinfo {author} {\bibfnamefont {P.}~\bibnamefont
  {Duval}},\ }\href@noop {} {\bibfield  {journal} {\bibinfo  {journal} {Int. J.
  Mater. Res.}\ }\textbf {\bibinfo {volume} {100}},\ \bibinfo {pages} {1433}
  (\bibinfo {year} {2009})}\BibitemShut {NoStop}%
\bibitem [{\citenamefont {Rodney}\ \emph {et~al.}(2011)\citenamefont {Rodney},
  \citenamefont {Tanguy},\ and\ \citenamefont
  {Vandembroucq}}]{msmse_19-083001-2011}%
  \BibitemOpen
  \bibfield  {author} {\bibinfo {author} {\bibfnamefont {D.}~\bibnamefont
  {Rodney}}, \bibinfo {author} {\bibfnamefont {A.}~\bibnamefont {Tanguy}}, \
  and\ \bibinfo {author} {\bibfnamefont {D.}~\bibnamefont {Vandembroucq}},\
  }\href@noop {} {\bibfield  {journal} {\bibinfo  {journal} {Modelling Simul.
  Mater. Sci. Eng.}\ }\textbf {\bibinfo {volume} {19}},\ \bibinfo {pages}
  {083001} (\bibinfo {year} {2011})}\BibitemShut {NoStop}%
\bibitem [{\citenamefont {Bulatov}\ and\ \citenamefont
  {Argon}(1999)}]{msmse_2-167-1994}%
  \BibitemOpen
  \bibfield  {author} {\bibinfo {author} {\bibfnamefont {V.~V.}\ \bibnamefont
  {Bulatov}}\ and\ \bibinfo {author} {\bibfnamefont {a.~S.}\ \bibnamefont
  {Argon}},\ }\href@noop {} {\bibfield  {journal} {\bibinfo  {journal}
  {Modelling Simul. Mater. Sci. Eng.}\ }\textbf {\bibinfo {volume} {2}},\
  \bibinfo {pages} {167} (\bibinfo {year} {1999})}\BibitemShut {NoStop}%
\bibitem [{\citenamefont {Picard}\ \emph {et~al.}(2004)\citenamefont {Picard},
  \citenamefont {Ajdari}, \citenamefont {Lequeux},\ and\ \citenamefont
  {Bocquet}}]{picard2004}%
  \BibitemOpen
  \bibfield  {author} {\bibinfo {author} {\bibfnamefont {G.}~\bibnamefont
  {Picard}}, \bibinfo {author} {\bibfnamefont {A.}~\bibnamefont {Ajdari}},
  \bibinfo {author} {\bibfnamefont {F.}~\bibnamefont {Lequeux}}, \ and\
  \bibinfo {author} {\bibfnamefont {L.}~\bibnamefont {Bocquet}},\ }\href@noop
  {} {\bibfield  {journal} {\bibinfo  {journal} {Eur. Phys. J. E.}\ }\textbf
  {\bibinfo {volume} {15}},\ \bibinfo {pages} {371} (\bibinfo {year}
  {2004})}\BibitemShut {NoStop}%
\bibitem [{\citenamefont {Martens}\ \emph {et~al.}(2011)\citenamefont
  {Martens}, \citenamefont {Bocquet},\ and\ \citenamefont
  {Barrat}}]{prl_106-156001-2011}%
  \BibitemOpen
  \bibfield  {author} {\bibinfo {author} {\bibfnamefont {K.}~\bibnamefont
  {Martens}}, \bibinfo {author} {\bibfnamefont {L.}~\bibnamefont {Bocquet}}, \
  and\ \bibinfo {author} {\bibfnamefont {J.-L.}\ \bibnamefont {Barrat}},\
  }\href@noop {} {\bibfield  {journal} {\bibinfo  {journal} {Phys. Rev. Lett.}\
  }\textbf {\bibinfo {volume} {106}},\ \bibinfo {pages} {156001} (\bibinfo
  {year} {2011})}\BibitemShut {NoStop}%
\bibitem [{\citenamefont {Nicolas}\ \emph {et~al.}(2014)\citenamefont
  {Nicolas}, \citenamefont {Martens}, \citenamefont {Bocquet},\ and\
  \citenamefont {Barrat}}]{Nicolas2014}%
  \BibitemOpen
  \bibfield  {author} {\bibinfo {author} {\bibfnamefont {A.}~\bibnamefont
  {Nicolas}}, \bibinfo {author} {\bibfnamefont {K.}~\bibnamefont {Martens}},
  \bibinfo {author} {\bibfnamefont {L.}~\bibnamefont {Bocquet}}, \ and\
  \bibinfo {author} {\bibfnamefont {J.-L.}\ \bibnamefont {Barrat}},\
  }\href@noop {} {\bibfield  {journal} {\bibinfo  {journal} {Soft Matter}\
  }\textbf {\bibinfo {volume} {10}},\ \bibinfo {pages} {4648} (\bibinfo {year}
  {2014})}\BibitemShut {NoStop}%
\bibitem [{\citenamefont {Fielding}(2014)}]{fielding2014}%
  \BibitemOpen
  \bibfield  {author} {\bibinfo {author} {\bibfnamefont {S.}~\bibnamefont
  {Fielding}},\ }\href@noop {} {\bibfield  {journal} {\bibinfo  {journal} {Rep.
  Prog. Phys.}\ }\textbf {\bibinfo {volume} {77}},\ \bibinfo {pages} {102601}
  (\bibinfo {year} {2014})}\BibitemShut {NoStop}%
\bibitem [{\citenamefont {Bouttes}\ and\ \citenamefont
  {Vandembroucq}(2013)}]{Bouttes2013}%
  \BibitemOpen
  \bibfield  {author} {\bibinfo {author} {\bibfnamefont {D.}~\bibnamefont
  {Bouttes}}\ and\ \bibinfo {author} {\bibfnamefont {D.}~\bibnamefont
  {Vandembroucq}},\ }\href@noop {} {\bibfield  {journal} {\bibinfo  {journal}
  {AIP Conference Proceedings}\ ,\ \bibinfo {pages} {481}} (\bibinfo {year}
  {2013})}\BibitemShut {NoStop}%
\bibitem [{\citenamefont {Homer}\ \emph {et~al.}(2010)\citenamefont {Homer},
  \citenamefont {Rodney},\ and\ \citenamefont {Schuh}}]{Homer2010}%
  \BibitemOpen
  \bibfield  {author} {\bibinfo {author} {\bibfnamefont {E.~R.}\ \bibnamefont
  {Homer}}, \bibinfo {author} {\bibfnamefont {D.}~\bibnamefont {Rodney}}, \
  and\ \bibinfo {author} {\bibfnamefont {C.~A.}\ \bibnamefont {Schuh}},\
  }\href@noop {} {\bibfield  {journal} {\bibinfo  {journal} {Phys. Rev. B}\
  }\textbf {\bibinfo {volume} {81}},\ \bibinfo {pages} {064204} (\bibinfo
  {year} {2010})}\BibitemShut {NoStop}%
\bibitem [{\citenamefont {Eshelby}(1957)}]{Eshelby1957}%
  \BibitemOpen
  \bibfield  {author} {\bibinfo {author} {\bibfnamefont {J.}~\bibnamefont
  {Eshelby}},\ }\href@noop {} {\bibfield  {journal} {\bibinfo  {journal} {Proc.
  R. Soc. London, Ser. A}\ }\textbf {\bibinfo {volume} {241}},\ \bibinfo
  {pages} {467} (\bibinfo {year} {1957})}\BibitemShut {NoStop}%
\bibitem [{\citenamefont {Daehn}(2001)}]{acma_49-2017-2001}%
  \BibitemOpen
  \bibfield  {author} {\bibinfo {author} {\bibfnamefont {G.~S.}\ \bibnamefont
  {Daehn}},\ }\href@noop {} {\bibfield  {journal} {\bibinfo  {journal} {Acta
  Mater.}\ }\textbf {\bibinfo {volume} {49}},\ \bibinfo {pages} {2017}
  (\bibinfo {year} {2001})}\BibitemShut {NoStop}%
\bibitem [{\citenamefont {Martens}\ \emph {et~al.}(2012)\citenamefont
  {Martens}, \citenamefont {Bocquet},\ and\ \citenamefont
  {Barrat}}]{Martens2012}%
  \BibitemOpen
  \bibfield  {author} {\bibinfo {author} {\bibfnamefont {K.}~\bibnamefont
  {Martens}}, \bibinfo {author} {\bibfnamefont {L.}~\bibnamefont {Bocquet}}, \
  and\ \bibinfo {author} {\bibfnamefont {J.-L.}\ \bibnamefont {Barrat}},\
  }\href@noop {} {\bibfield  {journal} {\bibinfo  {journal} {Soft Matter}\
  }\textbf {\bibinfo {volume} {8}},\ \bibinfo {pages} {4197} (\bibinfo {year}
  {2012})}\BibitemShut {NoStop}%
\bibitem [{\citenamefont {Hasan}\ and\ \citenamefont
  {Boyce}(1995)}]{Hasan1995}%
  \BibitemOpen
  \bibfield  {author} {\bibinfo {author} {\bibfnamefont {O.~A.}\ \bibnamefont
  {Hasan}}\ and\ \bibinfo {author} {\bibfnamefont {M.~C.}\ \bibnamefont
  {Boyce}},\ }\href@noop {} {\bibfield  {journal} {\bibinfo  {journal} {Polym.
  Eng. Sci.}\ }\textbf {\bibinfo {volume} {35}},\ \bibinfo {pages} {331–344}
  (\bibinfo {year} {1995})}\BibitemShut {NoStop}%
\bibitem [{\citenamefont {Shirmacher}\ \emph {et~al.}(2015)\citenamefont
  {Shirmacher}, \citenamefont {Ruocco},\ and\ \citenamefont
  {Mazzone}}]{Schirmacher2015}%
  \BibitemOpen
  \bibfield  {author} {\bibinfo {author} {\bibfnamefont {W.}~\bibnamefont
  {Shirmacher}}, \bibinfo {author} {\bibfnamefont {G.}~\bibnamefont {Ruocco}},
  \ and\ \bibinfo {author} {\bibfnamefont {V.}~\bibnamefont {Mazzone}},\
  }\href@noop {} {\bibfield  {journal} {\bibinfo  {journal} {Phys. Rev. Lett.}\
  }\textbf {\bibinfo {volume} {115}},\ \bibinfo {pages} {015901} (\bibinfo
  {year} {2015})}\BibitemShut {NoStop}%
\bibitem [{\citenamefont {Xia}\ and\ \citenamefont {Wolynes}(2001)}]{Xia2001}%
  \BibitemOpen
  \bibfield  {author} {\bibinfo {author} {\bibfnamefont {X.}~\bibnamefont
  {Xia}}\ and\ \bibinfo {author} {\bibfnamefont {P.~G.}\ \bibnamefont
  {Wolynes}},\ }\href@noop {} {\bibfield  {journal} {\bibinfo  {journal} {Phys.
  Rev. Lett.}\ }\textbf {\bibinfo {volume} {86}},\ \bibinfo {pages} {5526}
  (\bibinfo {year} {2001})}\BibitemShut {NoStop}%
\bibitem [{\citenamefont {Monthus}\ and\ \citenamefont
  {Bouchaud}(1996)}]{Bouchaud1996}%
  \BibitemOpen
  \bibfield  {author} {\bibinfo {author} {\bibfnamefont {C.}~\bibnamefont
  {Monthus}}\ and\ \bibinfo {author} {\bibfnamefont {J.-P.}\ \bibnamefont
  {Bouchaud}},\ }\href@noop {} {\bibfield  {journal} {\bibinfo  {journal} {J.
  Phys. A: Math. Gen.}\ }\textbf {\bibinfo {volume} {29}},\ \bibinfo {pages}
  {3847} (\bibinfo {year} {1996})}\BibitemShut {NoStop}%
\bibitem [{\citenamefont {Bortz}\ \emph {et~al.}(1975)\citenamefont {Bortz},
  \citenamefont {Kalos},\ and\ \citenamefont {Lebowitz}}]{Bortz1975}%
  \BibitemOpen
  \bibfield  {author} {\bibinfo {author} {\bibfnamefont {A.~B.}\ \bibnamefont
  {Bortz}}, \bibinfo {author} {\bibfnamefont {M.~H.}\ \bibnamefont {Kalos}}, \
  and\ \bibinfo {author} {\bibfnamefont {J.~L.}\ \bibnamefont {Lebowitz}},\
  }\href@noop {} {\bibfield  {journal} {\bibinfo  {journal} {J. Comput. Phys.}\
  }\textbf {\bibinfo {volume} {17}},\ \bibinfo {pages} {10} (\bibinfo {year}
  {1975})}\BibitemShut {NoStop}%
\bibitem [{\citenamefont {Fielding}\ \emph {et~al.}(2000)\citenamefont
  {Fielding}, \citenamefont {Sollich},\ and\ \citenamefont
  {Cates}}]{Fielding2000}%
  \BibitemOpen
  \bibfield  {author} {\bibinfo {author} {\bibfnamefont {S.~M.}\ \bibnamefont
  {Fielding}}, \bibinfo {author} {\bibfnamefont {P.}~\bibnamefont {Sollich}}, \
  and\ \bibinfo {author} {\bibfnamefont {M.~E.}\ \bibnamefont {Cates}},\
  }\href@noop {} {\bibfield  {journal} {\bibinfo  {journal} {J. Rheol}\
  }\textbf {\bibinfo {volume} {44}},\ \bibinfo {pages} {323} (\bibinfo {year}
  {2000})}\BibitemShut {NoStop}%
\bibitem [{\citenamefont {Sollich}(1998)}]{pre_58-738-1998}%
  \BibitemOpen
  \bibfield  {author} {\bibinfo {author} {\bibfnamefont {P.}~\bibnamefont
  {Sollich}},\ }\href@noop {} {\bibfield  {journal} {\bibinfo  {journal} {Phys.
  Rev. E}\ }\textbf {\bibinfo {volume} {58}},\ \bibinfo {pages} {738} (\bibinfo
  {year} {1998})},\ \Eprint {http://arxiv.org/abs/9712001} {arXiv:9712001
  [cond-mat]} \BibitemShut {NoStop}%
\bibitem [{\citenamefont {Hidalgo}\ \emph {et~al.}(2002)\citenamefont
  {Hidalgo}, \citenamefont {Kun},\ and\ \citenamefont
  {Herrmann}}]{pre_65-032502-2002}%
  \BibitemOpen
  \bibfield  {author} {\bibinfo {author} {\bibfnamefont {R.}~\bibnamefont
  {Hidalgo}}, \bibinfo {author} {\bibfnamefont {F.}~\bibnamefont {Kun}}, \ and\
  \bibinfo {author} {\bibfnamefont {H.}~\bibnamefont {Herrmann}},\ }\href@noop
  {} {\bibfield  {journal} {\bibinfo  {journal} {Phys. Rev. E}\ }\textbf
  {\bibinfo {volume} {65}},\ \bibinfo {pages} {032502} (\bibinfo {year}
  {2002})}\BibitemShut {NoStop}%
\bibitem [{\citenamefont {Pradhan}\ \emph {et~al.}(2010)\citenamefont
  {Pradhan}, \citenamefont {Hansen},\ and\ \citenamefont
  {Chakrabarti}}]{rmp_82-499-2010}%
  \BibitemOpen
  \bibfield  {author} {\bibinfo {author} {\bibfnamefont {S.}~\bibnamefont
  {Pradhan}}, \bibinfo {author} {\bibfnamefont {A.}~\bibnamefont {Hansen}}, \
  and\ \bibinfo {author} {\bibfnamefont {B.~K.}\ \bibnamefont {Chakrabarti}},\
  }\href@noop {} {\bibfield  {journal} {\bibinfo  {journal} {Rev. Mod. Phys.}\
  }\textbf {\bibinfo {volume} {82}},\ \bibinfo {pages} {499} (\bibinfo {year}
  {2010})}\BibitemShut {NoStop}%
\bibitem [{\citenamefont {Findley}\ \emph {et~al.}(1978)\citenamefont
  {Findley}, \citenamefont {Lai},\ and\ \citenamefont
  {Onaran}}]{book_flo-CreepRelaxViscoelastic}%
  \BibitemOpen
  \bibfield  {author} {\bibinfo {author} {\bibfnamefont {W.}~\bibnamefont
  {Findley}}, \bibinfo {author} {\bibfnamefont {J.}~\bibnamefont {Lai}}, \ and\
  \bibinfo {author} {\bibfnamefont {K.}~\bibnamefont {Onaran}},\ }\href@noop {}
  {\emph {\bibinfo {title} {{Creep and relaxation of nonlinear viscoelastic
  materials}}}}\ (\bibinfo  {publisher} {Dover},\ \bibinfo {year}
  {1978})\BibitemShut {NoStop}%
\bibitem [{\citenamefont {Feller}(1971)}]{book_Feller-IntroProbTheo2}%
  \BibitemOpen
  \bibfield  {author} {\bibinfo {author} {\bibfnamefont {W.}~\bibnamefont
  {Feller}},\ }\href@noop {} {\emph {\bibinfo {title} {{An introduction to
  probability theory and its applications, Volume II}}}},\ \bibinfo {edition}
  {2nd}\ ed.\ (\bibinfo  {publisher} {John Wiley \& Sons},\ \bibinfo {year}
  {1971})\BibitemShut {NoStop}%
\bibitem [{\citenamefont {Gueguen}\ \emph {et~al.}(2015)\citenamefont
  {Gueguen}, \citenamefont {Keryvin}, \citenamefont {Rouxel}, \citenamefont
  {Le-Fur}, \citenamefont {Orain}, \citenamefont {Bureau}, \citenamefont
  {Boussard-Pl\'edel},\ and\ \citenamefont {Sangleboeuf}}]{Gueguen2015}%
  \BibitemOpen
  \bibfield  {author} {\bibinfo {author} {\bibfnamefont {Y.}~\bibnamefont
  {Gueguen}}, \bibinfo {author} {\bibfnamefont {V.}~\bibnamefont {Keryvin}},
  \bibinfo {author} {\bibfnamefont {T.}~\bibnamefont {Rouxel}}, \bibinfo
  {author} {\bibfnamefont {M.}~\bibnamefont {Le-Fur}}, \bibinfo {author}
  {\bibfnamefont {H.}~\bibnamefont {Orain}}, \bibinfo {author} {\bibfnamefont
  {B.}~\bibnamefont {Bureau}}, \bibinfo {author} {\bibfnamefont
  {C.}~\bibnamefont {Boussard-Pl\'edel}}, \ and\ \bibinfo {author}
  {\bibfnamefont {J.-C.}\ \bibnamefont {Sangleboeuf}},\ }\href@noop {}
  {\bibfield  {journal} {\bibinfo  {journal} {Mechanics of Materials}\ }\textbf
  {\bibinfo {volume} {85}},\ \bibinfo {pages} {47} (\bibinfo {year}
  {2015})}\BibitemShut {NoStop}%
\bibitem [{\citenamefont {Dahmen}\ \emph {et~al.}(2009)\citenamefont {Dahmen},
  \citenamefont {Ben-Zion},\ and\ \citenamefont {Uhl}}]{prl_102-175501-2009}%
  \BibitemOpen
  \bibfield  {author} {\bibinfo {author} {\bibfnamefont {K.}~\bibnamefont
  {Dahmen}}, \bibinfo {author} {\bibfnamefont {Y.}~\bibnamefont {Ben-Zion}}, \
  and\ \bibinfo {author} {\bibfnamefont {J.}~\bibnamefont {Uhl}},\ }\href@noop
  {} {\bibfield  {journal} {\bibinfo  {journal} {Phys. Rev. Lett.}\ }\textbf
  {\bibinfo {volume} {102}},\ \bibinfo {pages} {175501} (\bibinfo {year}
  {2009})}\BibitemShut {NoStop}%
\bibitem [{\citenamefont {Fielding}\ \emph {et~al.}(2009)\citenamefont
  {Fielding}, \citenamefont {Cates},\ and\ \citenamefont
  {Sollich}}]{Fielding2009}%
  \BibitemOpen
  \bibfield  {author} {\bibinfo {author} {\bibfnamefont {S.~M.}\ \bibnamefont
  {Fielding}}, \bibinfo {author} {\bibfnamefont {M.~E.}\ \bibnamefont {Cates}},
  \ and\ \bibinfo {author} {\bibfnamefont {P.}~\bibnamefont {Sollich}},\
  }\href@noop {} {\bibfield  {journal} {\bibinfo  {journal} {Soft Matter}\
  }\textbf {\bibinfo {volume} {5}},\ \bibinfo {pages} {2378} (\bibinfo {year}
  {2009})}\BibitemShut {NoStop}%
\bibitem [{\citenamefont {Divoux}\ \emph {et~al.}(2010)\citenamefont {Divoux},
  \citenamefont {Tamarii}, \citenamefont {Barentin},\ and\ \citenamefont
  {Manneville}}]{prl_104-208301-2010}%
  \BibitemOpen
  \bibfield  {author} {\bibinfo {author} {\bibfnamefont {T.}~\bibnamefont
  {Divoux}}, \bibinfo {author} {\bibfnamefont {D.}~\bibnamefont {Tamarii}},
  \bibinfo {author} {\bibfnamefont {C.}~\bibnamefont {Barentin}}, \ and\
  \bibinfo {author} {\bibfnamefont {S.}~\bibnamefont {Manneville}},\
  }\href@noop {} {\bibfield  {journal} {\bibinfo  {journal} {Phys. Rev. Lett.}\
  }\textbf {\bibinfo {volume} {104}},\ \bibinfo {pages} {208301} (\bibinfo
  {year} {2010})}\BibitemShut {NoStop}%
\bibitem [{\citenamefont {Divoux}\ \emph {et~al.}(2012)\citenamefont {Divoux},
  \citenamefont {Tamarii}, \citenamefont {Barentin}, \citenamefont {Teitel},\
  and\ \citenamefont {Manneville}}]{sm_8-4151-2012}%
  \BibitemOpen
  \bibfield  {author} {\bibinfo {author} {\bibfnamefont {T.}~\bibnamefont
  {Divoux}}, \bibinfo {author} {\bibfnamefont {D.}~\bibnamefont {Tamarii}},
  \bibinfo {author} {\bibfnamefont {C.}~\bibnamefont {Barentin}}, \bibinfo
  {author} {\bibfnamefont {S.}~\bibnamefont {Teitel}}, \ and\ \bibinfo {author}
  {\bibfnamefont {S.}~\bibnamefont {Manneville}},\ }\href@noop {} {\bibfield
  {journal} {\bibinfo  {journal} {Soft Matter}\ }\textbf {\bibinfo {volume}
  {8}},\ \bibinfo {pages} {4151} (\bibinfo {year} {2012})}\BibitemShut
  {NoStop}%
\bibitem [{\citenamefont {Gopalakrishnan}\ and\ \citenamefont
  {Zukoski}(2007)}]{jrheo_51-623-2007}%
  \BibitemOpen
  \bibfield  {author} {\bibinfo {author} {\bibfnamefont {V.}~\bibnamefont
  {Gopalakrishnan}}\ and\ \bibinfo {author} {\bibfnamefont {C.~F.}\
  \bibnamefont {Zukoski}},\ }\href@noop {} {\bibfield  {journal} {\bibinfo
  {journal} {J. Rheol.}\ }\textbf {\bibinfo {volume} {51}},\ \bibinfo {pages}
  {623} (\bibinfo {year} {2007})}\BibitemShut {NoStop}%
\bibitem [{\citenamefont {Lindstr{\"{o}}m}\ \emph {et~al.}(2012)\citenamefont
  {Lindstr{\"{o}}m}, \citenamefont {Kodger}, \citenamefont {Sprakel},\ and\
  \citenamefont {Weitz}}]{sm_8-3657-2012}%
  \BibitemOpen
  \bibfield  {author} {\bibinfo {author} {\bibfnamefont {S.~B.}\ \bibnamefont
  {Lindstr{\"{o}}m}}, \bibinfo {author} {\bibfnamefont {T.~E.}\ \bibnamefont
  {Kodger}}, \bibinfo {author} {\bibfnamefont {J.}~\bibnamefont {Sprakel}}, \
  and\ \bibinfo {author} {\bibfnamefont {D.~A.}\ \bibnamefont {Weitz}},\
  }\href@noop {} {\bibfield  {journal} {\bibinfo  {journal} {Soft Matter}\
  }\textbf {\bibinfo {volume} {8}},\ \bibinfo {pages} {3657} (\bibinfo {year}
  {2012})}\BibitemShut {NoStop}%
\end{thebibliography}%
